\documentclass[conference]{IEEEtran}
\usepackage{cite}

\ifCLASSINFOpdf
   \usepackage[pdftex]{graphicx}
   \DeclareGraphicsExtensions{.pdf,.jpeg,.png}
\else
\fi
\usepackage{amsmath}
\usepackage{array}

  \usepackage[caption=false,font=footnotesize]{subfig}

\usepackage{stfloats}
\usepackage{url}

\begin{document}
%
\title{Microwave Tube Fault-Current Model for\\Design of Crowbar Protection}
 \author{\IEEEauthorblockN{Subhash Joshi T.G.}
 \IEEEauthorblockA{Member, IEEE\\Power Electronics Group\\
 Centre for Development of Advanced Computing\\
 Thiruvananthapuram-695033, India\\
 Email:subhashj@cdac.in}
 \and
 \IEEEauthorblockN{Vinod John}
 \IEEEauthorblockA{Senior Member, IEEE\\Department of Electrical Engineering\\
 Indian Institute of Science\\
 Bangalore-560012, India\\
 Email:vjohn@iisc.ac.in}
 }

%


\maketitle

\begin{abstract}
Many applications that use high energy plasma are realized using Microwave tubes (MWT) that operate at peak power in the range of hundreds of MW and frequency in GHz. One failure mode of the MWT is due to the excess energy in the tube during internal arcing events. Crowbar is used to protect the MWT by diverting the energy during fault. To compute the energy released into the MWT, the dc fault current model and the MWT model are essential. An equivalent fuse wire model is utilized for the MWT for the crowbar applications. The paper proposes a model for the dc fault current, the analysis for which is based on Joules Integral energy concept. The model provides flexibility to choose a range of practically observed reactance to resistance ratio ($X{/}R$) of transformer and also allows the use of a range of dc current limiting resistances that are utilized in the High Voltage (HV) power supply circuits in Microwave applications. The non-linearity of the system due to the multipulse diode rectifier is also considered by introducing a correction factor in the model. This paper shows that the same correction factor can be applied for both dc side parallel and series connected rectifier circuits. Both dc fault current and MWT models are verified experimentally. Using the model a $10kV$, $1kA$ crowbar is built to limit the energy in MWT below $10J$.
\end{abstract}

%
\IEEEpeerreviewmaketitle

\section{Introduction}
\begin{figure}[!t]
  \centering
  \subfloat[]{\includegraphics[keepaspectratio,scale=.67]{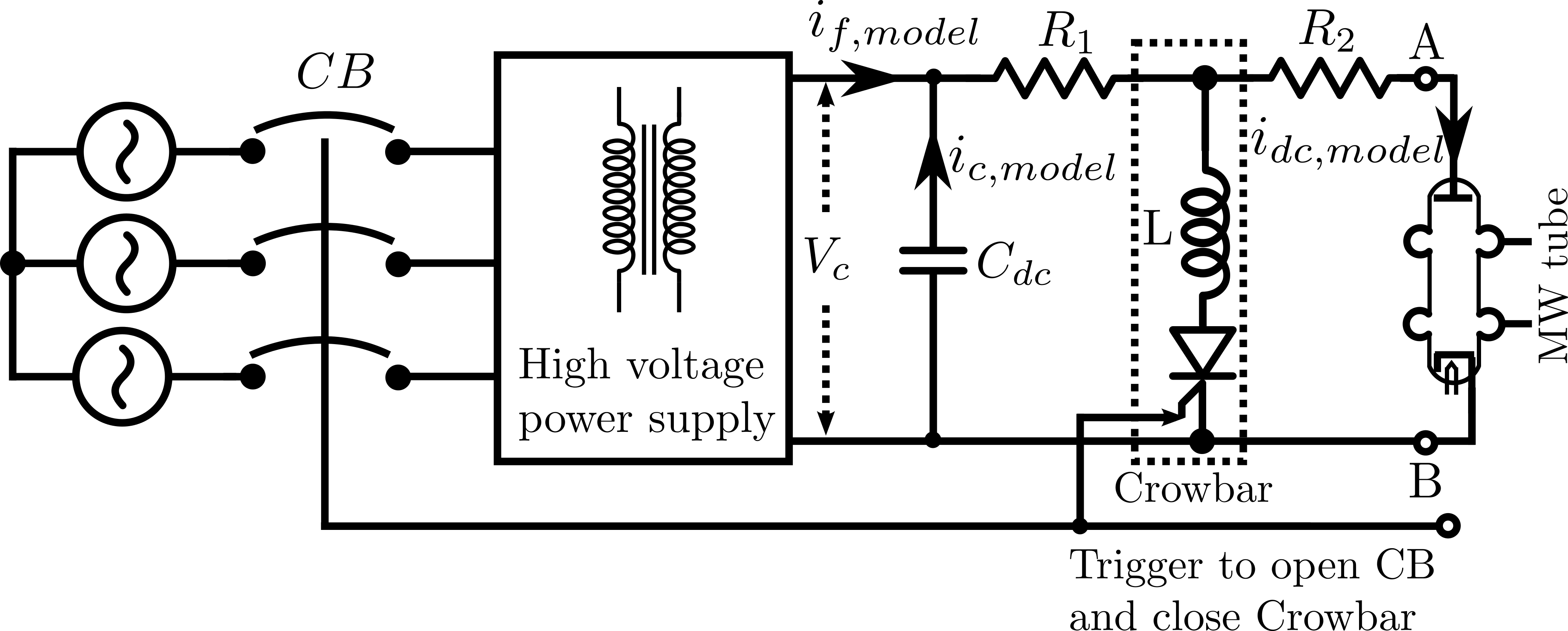}}\hspace{.12in}
  \subfloat[]{\includegraphics[keepaspectratio,scale=.67]{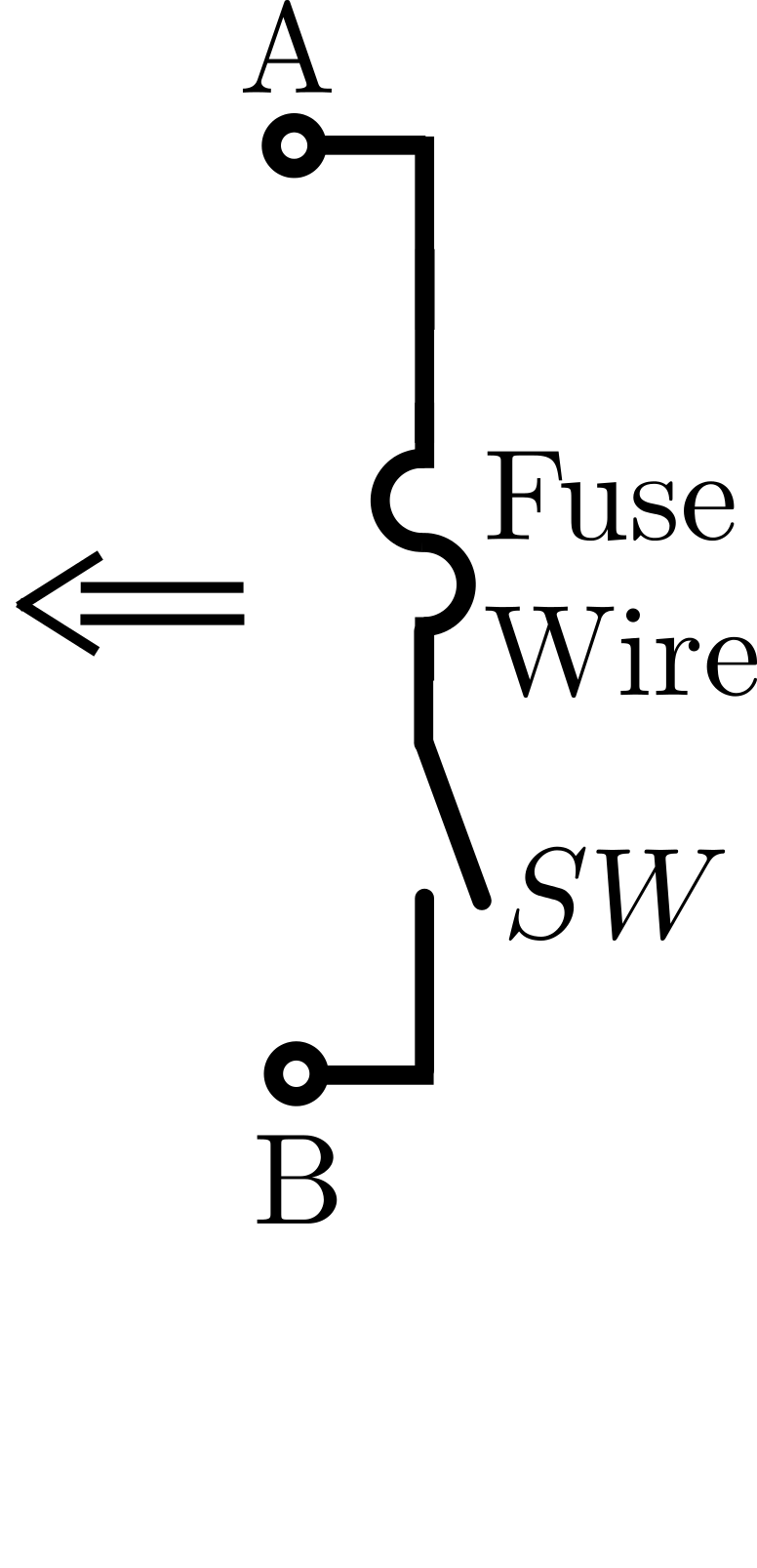}}
  \caption{Circuit network used for crowbar analysis (a) HV power supply with MWT and crowbar (b) MWT is replaced with fuse wire for wire survivability test.}
  \label{Fig1}
\end{figure}

%
%
%
%
\IEEEPARstart{P}{lasma} state of matter is widely used in many areas such as biomedical, material processing, electronics, textiles, space and defence~\cite{ref1,ref2,ref3,ref4,ref5,ref6,ref7}. 
The MWT manufacturers specify a limit on the energy that can be released into the tube during internal arc fault. For many tubes this limit is of the order of $10J$. During fault, if the energy accumulated inside the tube exceeds the specified value then the MWT becomes irreparable~\cite{ref11}. Conventional HV power supply built using mains frequency rectifier feeding power to MWTs, will have stored energy comparable to the energy limit of MWTs. Along with this, the slow operation of the circuit breaker (CB) in isolating the power supply from the grid also results in accumulation of additional energy into the tube, resulting in fault energy being above the specified limit. Hence fast acting protection is essential to keep the fault energy below the specified limit. 
This is achieved by turning on the 
crowbar connected in shunt with MWT, providing an alternative path for the flow of energy as shown in Fig. \ref{Fig1}(a)~\cite{ref15}. 
The energy released into the MWT is decided by the speed of operation of crowbar. 
The energy in MWT can be estimated if the model of dc fault current as well as the model of MWT are known. Such models are useful for the selection of speed of operation of crowbar. The model of fault current also helps in the design of various other components of crowbar such as selection of thyristor, its gate driver and design of the inductor.

Models for a rectifier are extensively discussed in literature. In \cite{ref8} dc fault current is estimated by assuming zero fault impedance and the model is described using multiple equations, each relevant for specific intervals. In \cite{ref9_} an averaged model is derived for the rectifier where the transients in the dc fault current are neglected. Model in \cite{ref10} can only be solved numerically due to its complexity. Due to the assumption of zero fault impedance most of the modelling methods are not suitable for crowbar applications, since during fault in the MWT the fault impedance is not zero, but is equal to $\left(R_1+R_2\right)$ as shown in Fig. \ref{Fig1}(a). Also, the value of $R_1$ and $R_2$ varies widely depending on the type of MWT and the rating of crowbar. Reference~\cite{ref11} describes a physical emulation of the MWT during fault conditions without detailing the mathematical model.

In this paper a dc fault current model based on the Joules Integral energy equivalence concept is proposed. 
Since the primary objective is to limit the energy accumulated in the MWT, the model based on energy concept is shown to lead to a more accurate and simple solution for the fault current and dissipated energy. The proposed model allows one to choose a range of practically observed $X{/}R$ ratios for the transformer. It also gives flexibility to choose various values of current limiting resistances $R_1$ and $R_2$ that are used with MWTs. The non-linearity of the rectifier system is also included by introducing a correction factor in the model. This correction factor is shown to be independent of ($X{/}R$) of transformer.

In~\cite{ref15_}, the dc fault current model is discussed for dc side is parallel connected rectifier system. In this paper the analysis is extended to dc side is series connected rectifier system also. From the analysis a generalized dc fault current model is derived which can be applied for both dc side series and parallel connected rectifier systems. The paper also shows that the correction factor required to the model for both dc side series and parallel connected circuits are identical. The dc fault current model for both series and parallel connection of ac-dc $12$ pulse rectifier bridges and MWT models are verified experimentally. A good match is observed between the results from the proposed analytical model and from the experimental hardware, with accuracy of better than 5\%.
\section{Proposed model for MWT during fault}
During internal arc, the MWT is emulated with a fuse wire made of copper
~\cite{ref11}\cite{ref12}. Hence for the performance evaluation of crowbar, in the wire survivability test the MWT is replaced with fuse wire of $10J$ and a HV switch $SW$ as shown in Fig. \ref{Fig1}(b). The fault is emulated by turning on $SW$
and this information is send to turn on crowbar and to open the \textit{CB}, which takes around $100ms$ to open
. During this test if the fuse wire survives without fusing, it is an indication that MWT will be protected by the crowbar during actual operation~\cite{ref12}. For the selection of crowbar the energy in fuse wire during the fault has to be evaluated, and hence a good model of fuse wire is essential.

The literature on fuse wire model with heat transfer equation classify the operation of fuse into $\left(i\right)$ pre-arcing phase 
and $\left(ii\right)$ arcing phase
~\cite{ref13}. For short fusing time where fuse wire carries a large current, the heat transfer equation can be solved only by considering thermal storage term in the energy balance equation \cite{ref13}\cite{ref14}. Since crowbar operation is limited only for the $100ms$ to open input \textit{CB}, a model for fuse wire is proposed by neglecting the heat transfer due to conduction, convection and radiation. Other assumptions applied considering short fusing time are $\left(i\right)$ uniform temperature rise $\left(ii\right)$ linear variation of electrical resistivity ($1/\sigma$) with temperature $\left(iii\right)$ skin effect; thermal expansion on length and area, oxidation of surface are neglected~\cite{ref14}. The model of fuse wire is also limited to pre-arc phase since for the acceptance of crowbar, the fuse wire has to be intact after wire survivability test. The incremental form of heat transfer equation for fuse wire having incremental temperature $\Delta T_f$ in an incremental time $\Delta t$ and carrying current $i_{fw}(t)$ is,
\begin{equation}\label{heat_eqn_modi}
i_{fw}^2(t)R_{fw}\Delta t=Al\rho~\mathcal{C}_p\Delta T_f
\end{equation}
where, $A$, $l$, $\rho$, $\mathcal{C}_p$ and $R_{fw}$ represents the area in $m^2$, length in $m$, mass density in $kg/m^3$, specific heat capacity in $J/kg^0C$ and resistance in $\Omega$ of fuse wire respectively. The resistance $R_{fw}$ is a function of fuse wire temperature $T_f$.
\subsection{Joules Integral and Area of fuse wire}\label{3.1.3}
Considering the variation of electrical conductivity, $\sigma$, with temperature, the fuse resistance is given by,
\begin{equation}\label{resistance}
R_{fw}=\dfrac{l}{\sigma\left(T_{f}\right) A}
\end{equation}
If $\sigma_o$ and $\alpha_o$ are the electrical conductivity in $S/m$ and temperature coefficient for electrical conductivity in $^0C^{-1}$ at temperature $T_o$, the electrical conductivity at any fuse temperature $T_f$ is~\cite{ref14},
\begin{equation}\label{conductivity}
\sigma(T_{f})=\dfrac{\sigma_o}{1+\alpha_0(T_{f}-T_o)}
\end{equation}
Substituting (\ref{conductivity}) in (\ref{resistance}), the resistance of fuse wire at time $t$ is given by,
\begin{equation}\label{res_t}
R_{fw}=\dfrac{1+\alpha_0(T_{f}-T_o)}{\sigma_o}\dfrac{l}{A}
\end{equation}
Substituting (\ref{conductivity}) and (\ref{resistance}) in (\ref{heat_eqn_modi}) and integrating over time $t$, the temperature of fuse wire at any time $t$ can be solved to be,
\begin{equation}\label{Tf}
T_{f}=\dfrac{1}{{\alpha_0}}\left(e^{\dfrac{\alpha_0}{A^2\rho~\mathit{C}_p\sigma_o}\int\limits_{0}^{t}i_{fw}^2(t)dt}-1\right)+T_o
\end{equation}
where $\int\limits_{0}^{t}i_{fw}^2(t)dt$ represents Joules Integral, $J_{I,t}$ of fuse wire at any time $t$.

If $T_m$ is the melting temperature of fuse wire and $J_{Im}$ is the Joules Integral at melting temperature, then from (\ref{Tf}) the area of fuse wire relates to $J_{Im}$ as,
\begin{equation}\label{JI_area}
A^2=K_{JI}J_{Im}
\end{equation}
where $K_{JI}$ is a constant that depends on the physical parameters of the fuse wire given by,
\begin{equation}\label{kji}
K_{JI}=\dfrac{\alpha_0}{\rho~\mathit{C}_p\sigma_o\ln\left[1+\alpha_0(T_m-T_o)\right]}
\end{equation}
From (\ref{JI_area}), the Joule Integral is independent of length of the fuse wire.
\begin{figure}[!t]
  \centering
\subfloat[]{\includegraphics[keepaspectratio,scale=0.75]{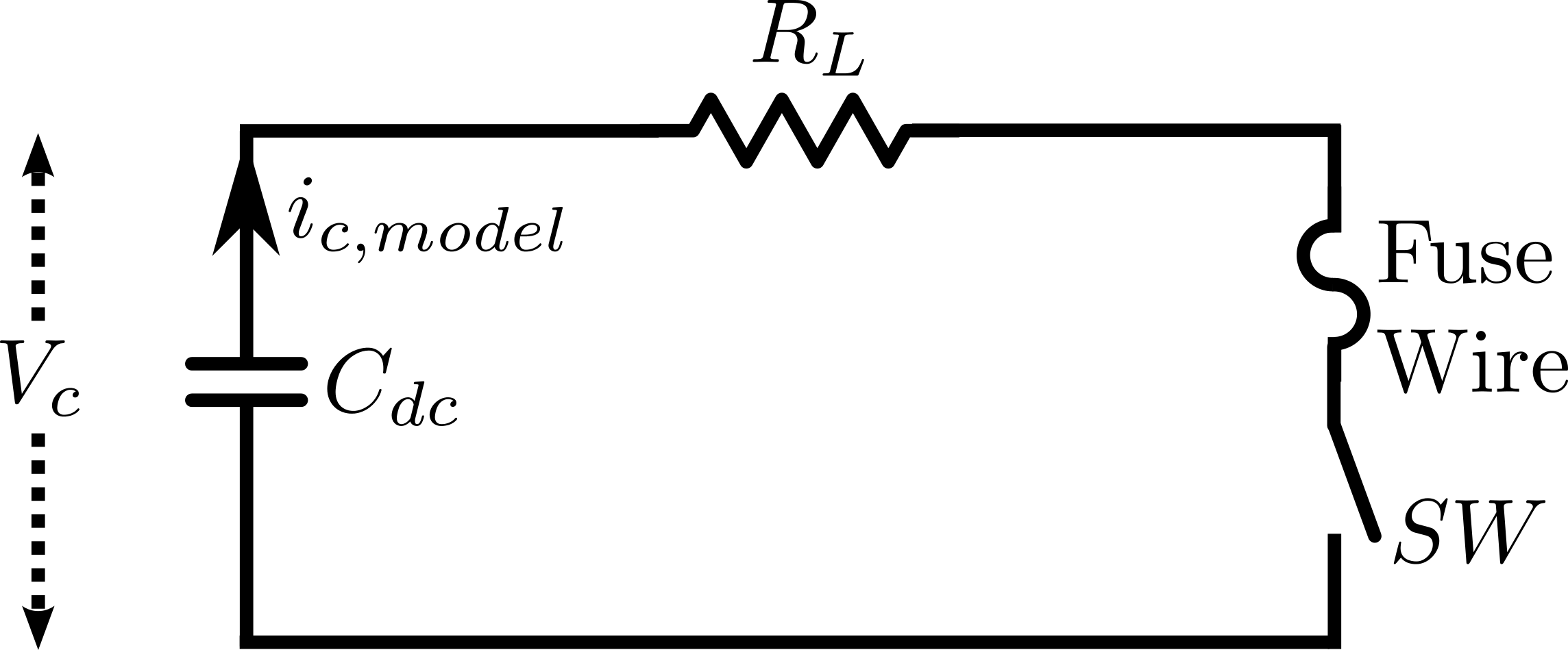}}\hspace{0.5in}
\subfloat[]{\includegraphics[keepaspectratio,scale=0.6]{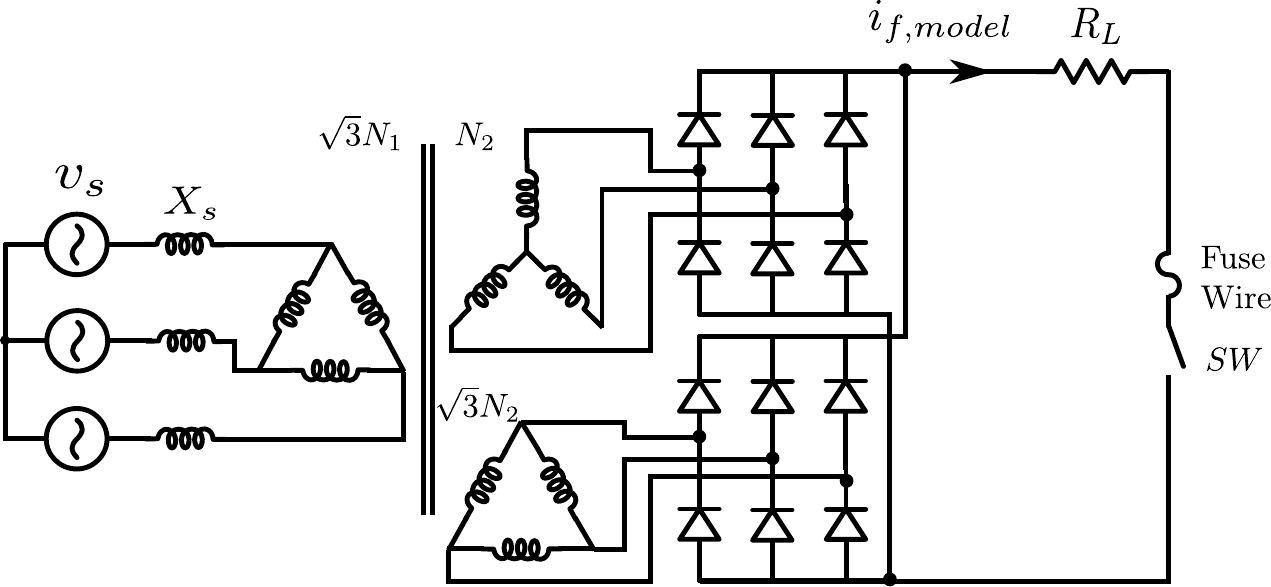}}
   \caption{Simplified equivalent circuit for (a) discharge current of dc output capacitor (b) follow-on current from input supply.} 
   \label{Fig2}
\end{figure}
\subsection{Energy and length of fuse wire}\label{3.1.4}
Substituting (\ref{Tf}) in (\ref{res_t}), the resistance of the fuse wire at any time $t$ is written as,
\begin{equation}\label{res_time}
R_{fw,t}=\dfrac{l}{\sigma_oA}e^{\dfrac{\alpha_o}{A^2\rho\mathit{C_p}\sigma_o}\int\limits_{0}^{t}i_{fw}^2(t)dt}
\end{equation}
Hence energy in the fuse wire is given by,
\begin{equation}\label{energy2}
E_{f,t}=\dfrac{l}{\sigma_oA}\int\limits_{0}^{t}i_{fw}^2(t)e^{\dfrac{\alpha_o}{A^2\rho\mathit{C_p}\sigma_o}\int\limits_{0}^{t}i_{fw}^2(t)dt}dt
\end{equation}
Since (\ref{energy2}) involves Joules Integral of current rather than the actual current, any profile of the current can be chosen. Hence to find the energy in the fuse wire at any time $t$, $i_{fw}(t)$ in (\ref{energy2}) is chosen as a dc current of magnitude $I_{dc}$. If fuse temperature reaches $T_m$ in time $t_m$, then by substituting (\ref{JI_area}) in (\ref{energy2}), the energy in the fuse wire at its melting temperature is given by,
\begin{equation}\label{energy4}
E_{fm}=\big(Al\big)\dfrac{\rho\mathit{C_p}}{\alpha_o}\left[e^{\dfrac{\alpha_o}{\rho\mathit{C_p}\sigma_oK_{JI}}}-1\right]
\end{equation}
Applying (\ref{kji}) in (\ref{energy4}), $E_{fm}$ is simplified to,
\begin{equation}\label{energy5}
E_{fm}=\big(Al\big)\rho\mathit{C_p}\left(T_m-T_o\right)
\end{equation}
From (\ref{energy5}), the product of area $A$ and length $l$ of a fuse wire is constant for a given energy and the length of the fuse wire can be computed. These relations are verified experimentally for the crowbar application.
\section{DC fault current model for crowbar application}
\subsection{Operation of conventional high voltage power supply}
In high energy high voltage plasma application, HV is built using $12$-pulse diode bridge rectifier 
to meet the required power level and to have smaller dc voltage ripple, 
where dc side of rectifiers are parallel connected as shown in Fig. \ref{Fig2}(b). The dc fault current initiated due to the internal arc of MWT consists of two components $\left(i\right)$ follow-on current from the input supply due to the delay in opening \textit{CB}, $i_{f,model}(t)$ $\left(ii\right)$ discharge current, $i_{c,model}(t)$, of dc output capacitor $C_{dc}$, shown in Fig. \ref{Fig1}(a). To minimize the energy accumulation into the MWT during internal arc, HV power supplies are characterized by lower $C_{dc}$ and higher ${X}{/}{R}$ ratio of transformer. With a transformer of short circuit power factor angle close to $\pi/2$, dc fault current peak occurs approximately at $10ms$ for a grid frequency of $50Hz$, where as the maximum dc capacitor discharge time constant during dc fault is less than $1ms$~\cite{ref8}. Hence these two components of fault current can be analysed independently before superimposing to obtain the total dc fault current $i_{dc,model}(t)$. The equivalent circuit relating to $i_{c,model}(t)$ and $i_{f,model}(t)$ are shown in Figs. \ref{Fig2}(a) and (b) respectively for a rectifier circuit of parallel connected dc output. Since the objective of the analysis is to find the model for $i_{dc,model}(t)$ due to the internal arc of MWT, crowbar is not considered in the analysis in the fault duration before it is triggered and hence $R_L=R_1+R_2$.
\subsection{Model for $i_{c,model}(t)$}
Since discharge current of $C_{dc}$ is only limited by $R_L$, this component decides the peak of $i_{dc,model}(t)$. If $C_{dc}$ is precharged to $V_c$, then the model for capacitor discharge current is,
\begin{equation}\label{ic}
i_{c,model}(t)=\dfrac{V_c}{R_L}e^{-\tfrac{t}{R_LC_{dc}}}
\end{equation}
\subsection{Model for $i_{f,model}(t)$ based on Joules Integral}
The principle of modelling of the follow-on fault current is based on the Joules Integral $J_I$. The maximum error in Joules Integral at any time $t$, $J_{I,t}$, by the proposed model and simulated fault current $i_{f,sim}(t)$ is expected to be within $5\%$ for engineering accuracy. The error in $J_{I,t}$ is defined as,
\begin{equation}\label{err}
\Delta J_{I\%,t}=\dfrac{\int\limits_{0}^{t}i^2_{f,sim}(t)dt-\int\limits_{0}^{t}i^2_{f,model}(t)dt}{\int\limits_{0}^{t}i^2_{f,sim}(t)dt}100 
\end{equation}
where, $\int\limits_{0}^{t}i^2_{f,sim}(t)dt$ is the Joules Integral evaluated from a detailed circuit simulation of the circuit in Fig.~\ref{Fig2}(b). The term $\int\limits_{0}^{t}i^2_{f,model}(t)dt$ is the analytical model derived below.
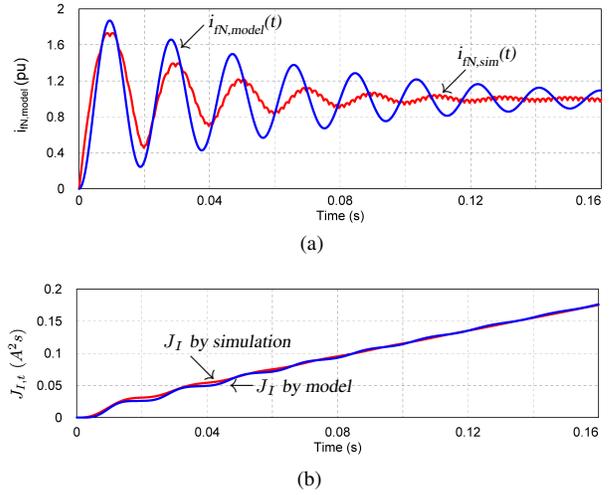
\begin{figure}[!t]
\centering
\subfloat[]{\includegraphics[keepaspectratio,scale=0.36]{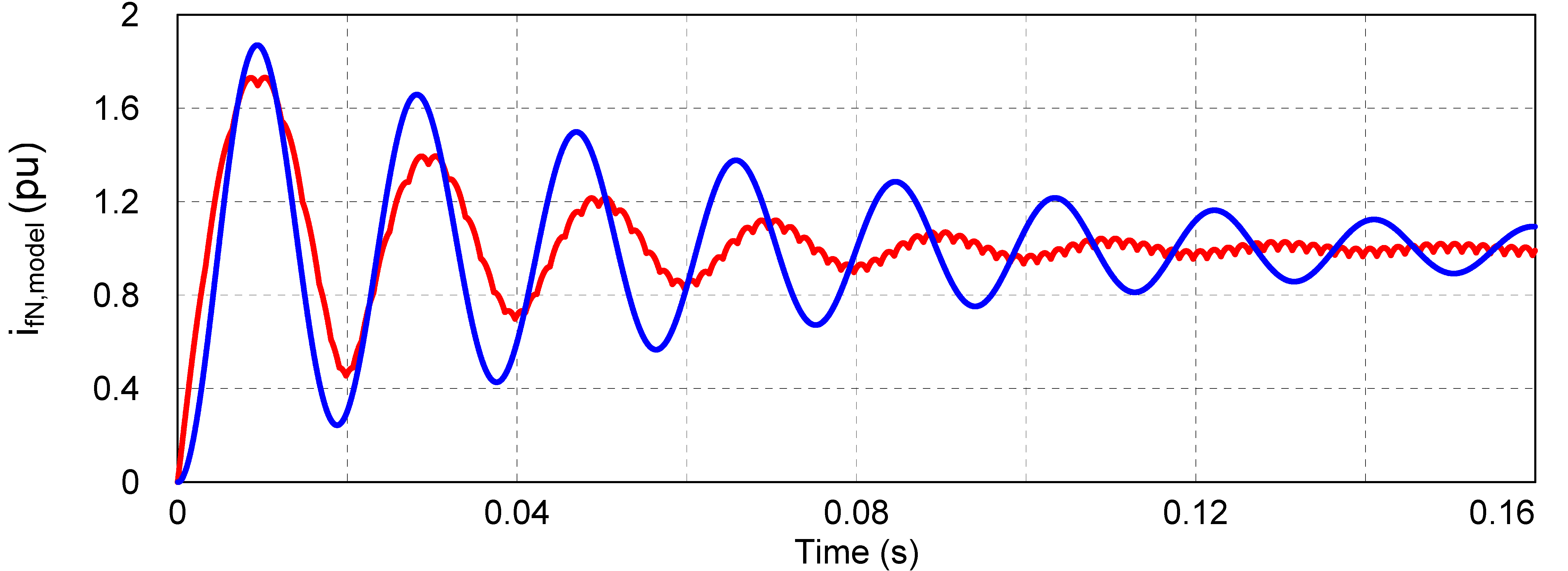}}
\put(-65,51){\textsl{$\swarrow$}}
\put(-59,62){\textsl{\scriptsize i\raisebox{-.8ex}{\tiny fN,sim}(t)}}
\put(-164,68){\textsl{$\swarrow$}}
\put(-153,74){\textsl{\scriptsize i\raisebox{-.8ex}{\tiny fN,model}(t)}}\hspace{0.025in}
\subfloat[]{\includegraphics[keepaspectratio,scale=0.36]{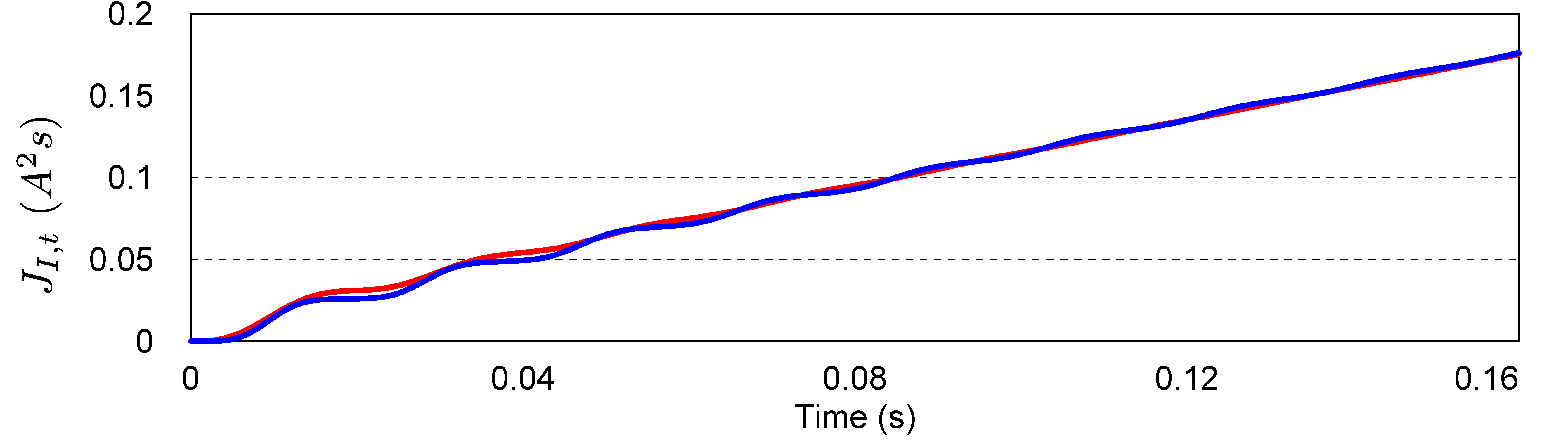}}
\put(-162,32){\textsl{$\searrow$}}
\put(-172,42){\textsl{\scriptsize $J_I$ by simulation}}
\put(-147,25){\textsl{$\leftarrow$}}
\put(-137,25){\textsl{\scriptsize $J_I$ by model}}\hspace{0.025in}
\caption{Fault current and Joules Integral. (a) Fault current using simulation and analytical model (b) Joules Integral using simulation and analytical model.}
\label{Fig3}
\end{figure}

The transformer considered for the HV power supply is having dual secondary where winding configuration is $\Delta/Y/\Delta$ with turns ratio of $N_1:N_2$ and $\sqrt{3}N_1:N_2$ between primary and delta secondary as well as primary and star secondary respectively as shown in Fig. \ref{Fig2}(b). The winding resistance and leakage reactance of star and delta secondary referred to primary are assumed to be equal and are given by $R_{sp}$ and $X_{l,sp}$ respectively. $R_{p\Delta}, X_{l,p\Delta}$ are the winding resistance and leakage reactance of primary respectively. The input source impedance $X_s$ referred to be a part of primary leakage reactance can be found by using reactive power equality. Then the equivalent resistance and reactance referred to primary for a secondary paralleled transformer by incorporating $X_s$ is,
\begin{equation}\label{Rtr}
R^{'}_{p}=R_{p\Delta}+\frac{R_{sp}}{2}\quad and \quad X^{'}_{lp}=X_{l,p\Delta}+\frac{X_{l,sp}}{2}+3X_s
\end{equation}
For a zero fault impedance at dc ($R_L=0$) and  primary line to line rms voltage of $E$, the steady state dc fault current which also chosen as the base value for current is,
\begin{equation}\label{ifaultdc}
I_{f,base}=\dfrac{\sqrt{2}E}{\sqrt{R_{p}^{'2}+X_{lp}^{'2}}}\dfrac{\sqrt{3}N_1}{N_2}k_{12}
\end{equation}
where, $k_{12}$ is equal to $0.9886$ and represents the ratio of average to peak value for $12$-pulse waveform \cite{ref8}.

If $X\_R_{trx}$ is the ratio of equivalent reactance $X^{'}_{lp}$ to equivalent resistance $R^{'}_{p}$, which includes transformer and input source impedance, then from simulation Fig. \ref{Fig3}(a) gives the normalized follow-on dc fault current $i_{fN,sim}(t)$ for $R_L=0$ and $X\_R_{trx}=10$. For maximum fault current peak, the fault initiation time is selected based on the power factor angle of the transformer \cite{ref8}. $i_{fN,sim}(t)$ given in Fig. \ref{Fig3}(a) resembles the unit step response of a second order system and hence the normalized model for $i_{f,model}(t)$ is chosen as,
\begin{equation}\label{2ndOrdereqn}
i_{fN,model}(t)=1-e^{-\delta t}\left(\cos\omega_dt+\dfrac{\delta}{\omega_d}\sin\omega_dt\right)
\end{equation}
where, $\omega_d$ and $\delta$ are the damped frequency of the oscillation and reciprocal of time constant for exponential decay respectively.

\begin{figure}[!t]
\centering
{\includegraphics[keepaspectratio,scale=0.34]{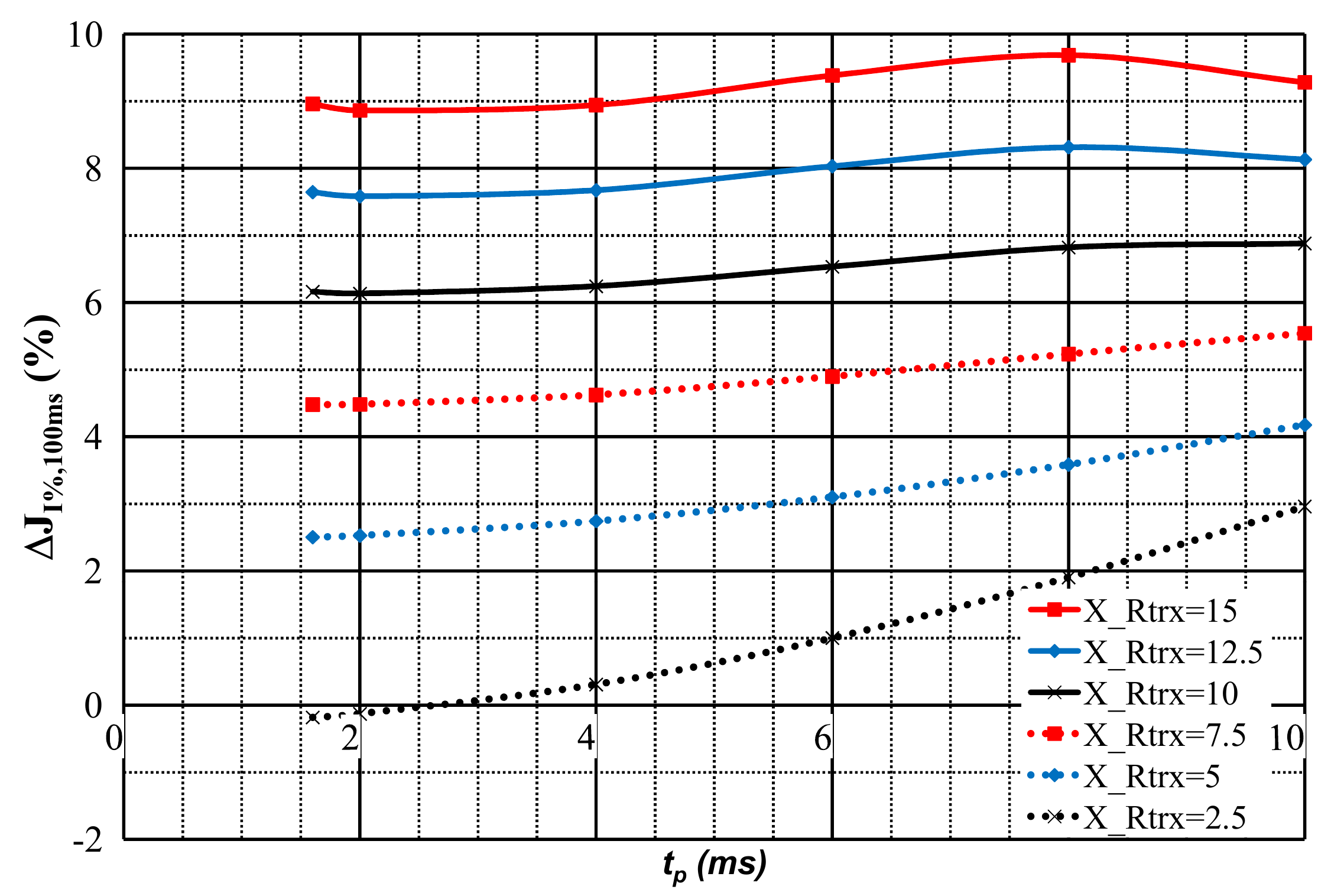}}
\caption{Percentage Joules Integral error for various time to peak fault currents $t_p\in\left[1.5ms,10ms\right]$ and transformer $X{/}R$ ratios $X{/}R\in\left[2.5,15\right]$.}
\label{Fig3c}
\end{figure}
If $t_p$ and $M_p$ are the time to reach its first peak and maximum overshoot respectively, they can be expressed in terms of $\omega_d$ and $\delta$ as,
\begin{equation}\label{wd}
t_p=\dfrac{\pi}{\omega_d}\qquad\qquad \text{and} \qquad\qquad 
M_p=e^{-\tfrac{\pi\delta}{\omega_d}}
\end{equation}

The parameters $\omega_d$ and $\delta$ are chosen such that $\Delta J_{I\%,t}$ computed with the model in (\ref{2ndOrdereqn}) and the follow-on current from simulation has maximum error less than $\pm 5\%$. Fig. \ref{Fig3}(a) compares $i_{fN,model}(t)$ and $i_{fN,sim}(t)$ for $\omega_d=334~rad/s$ and $\delta=14.8~s^{-1}$. Fig. \ref{Fig3}(b) shows that $J_{I,t}$ due to the empirical model in (\ref{2ndOrdereqn}) as well as the time domain simulated follow-on dc fault current are within $5\%$ error. The follow-on dc fault current from its normalized model in (\ref{2ndOrdereqn}) is given by,
\begin{equation}\label{if}
i_{f,model}(t)=I_{f,base}~i_{fN,model}(t)
\end{equation}
In practice the dc side fault impedance are not zero due to resistances $R_1$ and $R_2$ and the $X{/}R$ ratio of transformer varies based on the power rating of the application. The influence of $X{/}R$ ratio of transformer and $R_L$ on $J_I$ are analysed in Section \ref{sectionIV}.
\begin{figure}[!t]
\centering
\includegraphics[keepaspectratio,scale=0.35]{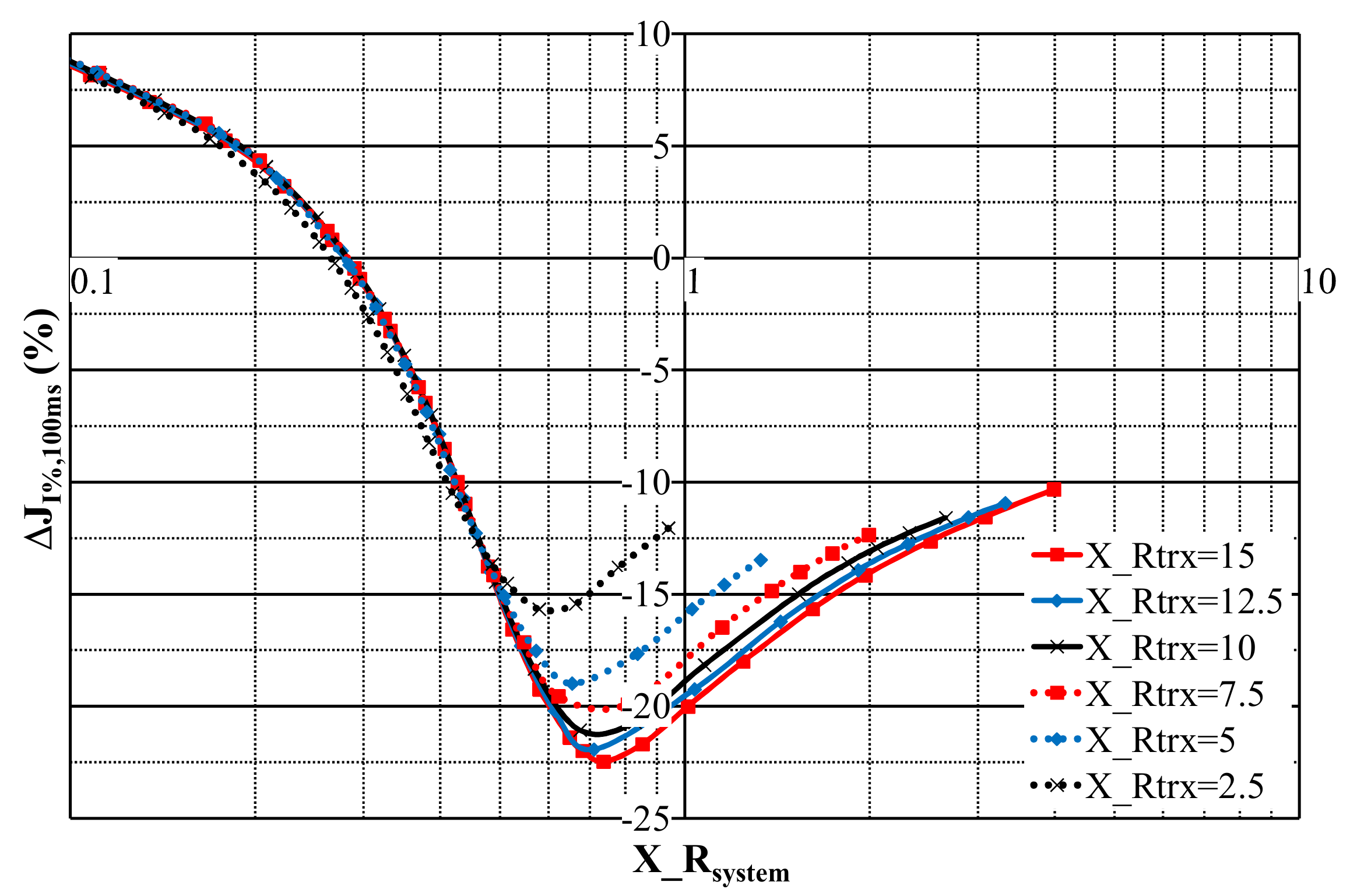}
\caption{$\Delta J_{I\%,100ms}$ for different $X\_R_{system}$ plotted for various $X\_R_{trx}$.}
\label{Fig4a}
\end{figure}
\section{Practical consideration}\label{sectionIV}
\subsection{Effect of transformer $X\_R_{trx}$ on $i_{f,model}(t)$}
Depending on the energy level of plasma required for the application, MWT of different ratings are used. This decides the rating of HV power supply and the rating of transformer that are used to build these HV power supplies. Hence in practice different $X{/}R$ ratio of transformers are found. The parameters $\omega_d$ and $\delta$ have dimensions of frequency and in this analysis they are influenced by $X^{'}_{lp}/R^{'}_{p}$ for a given operational frequency $\omega$. The transients that appear in the $i_{f,sim}(t)$ in Fig. \ref{Fig3}(a) are due the transients in the transformer input current and its exponential decay time constant is given by $X\_R_{trx}/\omega$. Hence for the model  $i_{f,model}(t)$, $\delta$ is chosen as,
\begin{equation}\label{delta}
\delta=\dfrac{\omega R^{'}_{p}}{X^{'}_{lp}}
\end{equation}
For $50Hz$ grid frequency $t_p$ is equal to $10ms$ for a transformer with short circuit power factor angle of $\pi/2$~\cite{ref8}. In the proposed model $t_p$ is chosen as a constant equal to $9.4ms$. However Fig. \ref{Fig3c} shows the evaluated $\Delta J_{I\%,t}$ at time $t=100ms$ for a given $X\_R_{trx}$ with $\delta$ in (\ref{delta}) for various values of $t_p$ between $1.5ms$ to $10ms$. The $\Delta J_{I\%,t}$ evaluated at $t=100ms$ is denoted by $\Delta J_{I\%,100ms}$. Fig. \ref{Fig3c} also shows the variation in $\Delta J_{I\%,100ms}$ for various $X\_R_{trx}$ found practically in ranges from $2.5$ to $15$. For any $X\_R_{trx}$ in Fig. \ref{Fig3c} the difference between the maximum and minimum value of $\Delta J_{I\%,100ms}$ is found to be less than $3.2\%$ when $t_p$ varies between $1.5ms$ to $10ms$. However, for any given $t_p$, $\Delta J_{I\%,100ms}$ can be observed to be increasing with $X\_R_{trx}$ giving a value close to $10\%$ for $X{/}R$ of $15$. It is desirable to reduce this error in the Joules Integral evaluated from the model.
\subsection{Influence of $R_L$ on $i_{f,model}(t)$}
The $R_L$ shown in Fig.~\ref{Fig2}(b) has a range of values based on the rating of the MWT used along with the HV power supply. The steady state dc fault current which is same as the base value of current varies with $R_L$. By approximating the transformer input current to fundamental, the resistance $R_L$ referred to primary, $R_{Lp}$, is evaluated by applying the active power equality in $R_{Lp}$ and $R_L$. If $I_{pw,rms}$ is the primary winding rms current, then by active power equality,
\begin{equation}\label{AP}
3I^{2}_{pw,rms}R_{Lp}=I^{2}_{f,base}R_L
\end{equation}
From (\ref{ifaultdc}), $I_{pw,rms}$ and $I_{f,base}$ are related to,
\begin{equation}\label{ipidc}
\dfrac{I_{f,base}}{I_{pw,rms}}=\sqrt{2}\dfrac{\sqrt{3}N_1}{N_2}k_{12}
\end{equation}
Substituting (\ref{ipidc}) in (\ref{AP}) gives,
\begin{equation}\label{rlp}
R_{Lp}=\dfrac{2}{3}{\left(\dfrac{\sqrt{3}N_1}{N_2}k_{12}\right)}^2R_L
\end{equation}
The resistance $R_{Lp}$ can be used to consider the impact of $R_1$ and $R_2$ in the fault current model by effectively transferring these resistances to the ac side while keeping the active power dissipation the same, as constrained by (\ref{AP}). This is done by modifying the $I_{f,base}$ in (\ref{ifaultdc}) and $\delta$ in (\ref{delta}) as,
\begin{equation}\label{ifaultdcM}
I_{f,base}=\dfrac{\sqrt{2}E}{\sqrt{\left(R^{'}_{p}+R_{Lp}\right)^{2}+X_{lp}^{'2}}}\dfrac{\sqrt{3}N_1}{N_2}k_{12}
\end{equation}
\begin{equation}\label{deltaM}
\delta=\dfrac{\omega \left(R^{'}_{p}+R_{Lp}\right)}{X^{'}_{lp}}
\end{equation}
In addition to $X\_R_{trx}$, introduction of $R_L$ requires the re-evaluation of $X/R$ for the system shown in Fig. \ref{Fig2}(b) as,
\begin{equation}\label{XRM}
X\_R_{system}=\dfrac{X^{'}_{lp}}{R^{'}_{p}+R_{Lp}}
\end{equation}
The modified model for the follow-on dc fault current by including the effect of $R_L$ is computed by substituting (\ref{2ndOrdereqn}) and (\ref{ifaultdcM}) in (\ref{if}) using the modified $\delta$ given in (\ref{deltaM}). For the analysis, $R_L$ is chosen in a practical range between $5\Omega$ and $300\Omega$, the respective $X\_R_{system}$ is computed for various values of $X\_R_{trx}$ of transformer. Fig. \ref{Fig4a} shows $\Delta J_{I\%,100ms}$ computed with the modified $i_{f,model}$ versus $X\_R_{system}$ for various values of $X\_R_{trx}$. From Fig. \ref{Fig4a}, $\Delta J_{I\%,100ms}$ is similar for different values of $X\_R_{trx}$ when $X\_R_{system}\le 0.5$ where as $\Delta J_{I\%,100ms}$ is similar for any $X\_R_{system}$ if $X\_R_{trx}\ge 7.5$. Fig. \ref{Fig4a} also indicate the large error introduced in $J_{I,100ms}$ by the assumption of equivalent power dissipation made in (\ref{AP}). The maximum positive and negative value of $\Delta J_{I\%,100ms}$ are found to be $8\%$ and $-22\%$ respectively. It is shown in the subsection below that a correction factor that is independent of $X\_R_{trx}$ can be used to bring the $\Delta J_{I\%,100ms}$ to a level less than $\pm 5\%$. 
\begin{figure}[!t]
\centering
\includegraphics[keepaspectratio,scale=0.34]{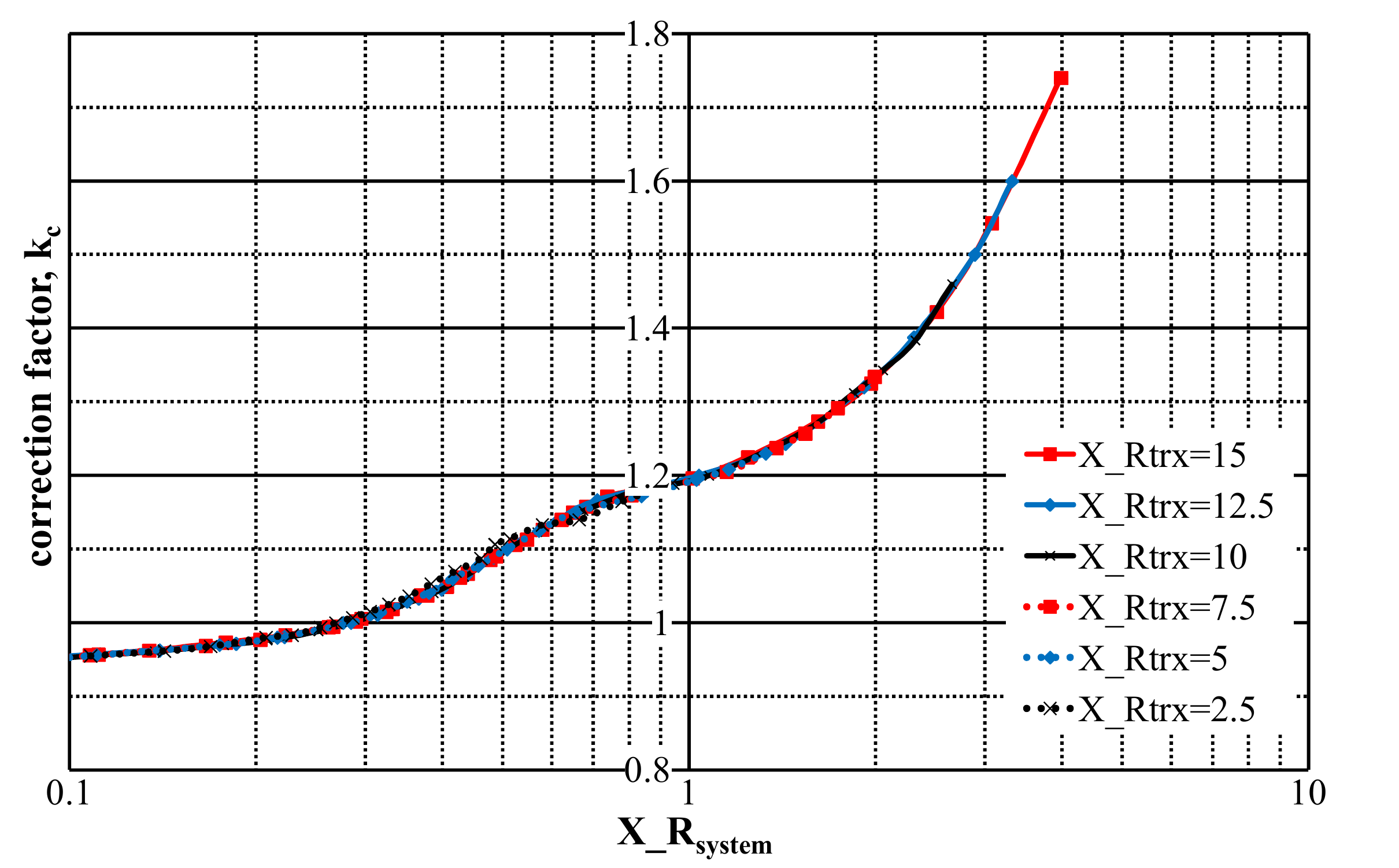}
\put(-181,8){\textsl{\normalsize $0.3$}}
\caption{Correction factor $k_c$ for different $X\_R_{system}$ plotted for various $X\_R_{trx}$.}
\label{Fig4b}
\end{figure}
\begin{figure}[!t]
\centering
\includegraphics[keepaspectratio,scale=0.34]{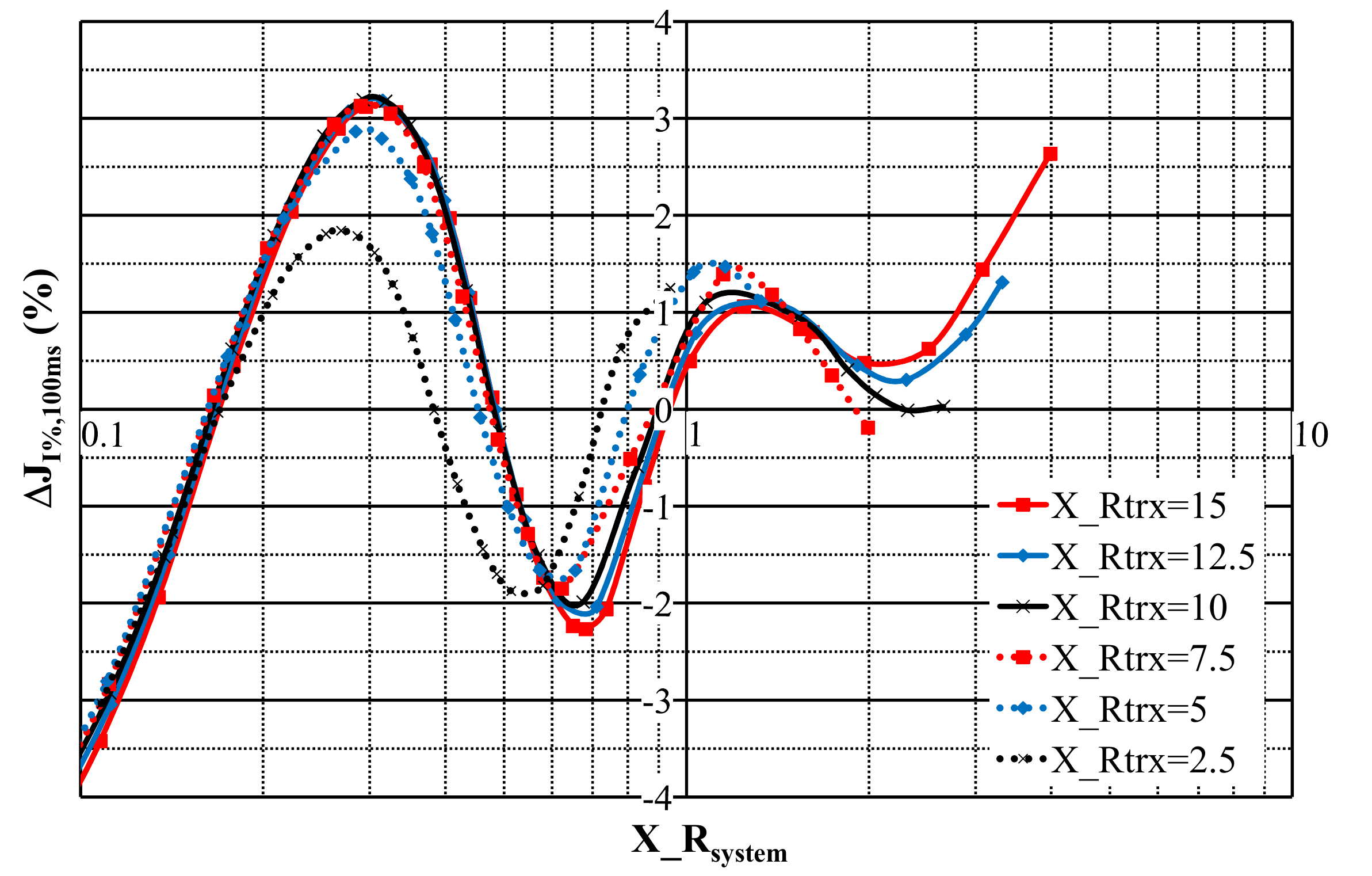}
\caption{$\Delta J_{I\%,100ms}$ for different $X\_R_{system}$ plotted for various $X\_R_{trx}$ after including $k_c$ factor for parallel connected dc output rectifier.}
\label{Fig4c}
\end{figure}
\subsection{Correction factor $k_c$}
From Fig. \ref{Fig3c} for a given $t_p$ the error $\Delta J_{I\%,100ms}$ increases with $X\_R_{trx}$ due to approximation in selecting $\delta$. Increase in $X\_R_{trx}$ implies a reduction in $\delta$ from (\ref{delta}). The error $\Delta J_{I\%,100ms}$ can be minimized by increasing $\delta$ appropriately as $X\_R_{trx}$ increases by introducing a correction factor to the resistance in (\ref{deltaM}). Hence, the correction factor required to compensate the approximation made for $\delta$ should increase with $X{/}R$ ratio. From Fig. \ref{Fig4a} $\Delta J_{I\%,100ms}$ is positive for $X\_R_{system}\le 0.3$ and beyond it is negative. Since this $\Delta J_{I\%,100ms}$ is due to the assumption of equivalent power dissipation in selecting $R_{Lp}$, the correction factor is applied to $R_{Lp}$ such that it is less than unity till $X\_R_{system}\le 0.3$ and greater than unity for  $X\_R_{system}>0.3$. Here also the correction factor should increase with the $X{/}R$ ratio. The correction factor required to reduce the $\Delta J_{I\%,100ms}$ in Fig. \ref{Fig3c} and the correction factor required to reduce the $\Delta J_{I\%,100ms}$ in Fig. \ref{Fig4a} show similar characteristics. This shows a consistency in the requirement for the correction factor $k_c$ that is applied to $R_{Lp}$ to compensate the equivalent power dissipation assumption in (\ref{AP}) and the assumption in selecting $\delta$ in (\ref{deltaM}). By applying $k_c$ the base value of current given in (\ref{ifaultdcM}) and $\delta$ in (\ref{deltaM}) are modified as,
\begin{equation}\label{ifaultdcMM}
I_{f,base}=\dfrac{\sqrt{2}E}{\sqrt{\left(R^{'}_{p}+k_cR_{Lp}\right)^{2}+X_{lp}^{'2}}}\dfrac{\sqrt{3}N_1}{N_2}k_{12}
\end{equation}
\begin{equation}\label{deltaMM}
\delta=\dfrac{\omega \left(R^{'}_{p}+k_cR_{Lp}\right)}{X^{'}_{lp}}
\end{equation}
Fig. \ref{Fig4b} shows the correction factor $k_c$ computed for various $X\_R_{system}$ to keep $\left|\Delta J_{I\%,100ms}\right|$ less than $5\%$ even when $X\_R_{trx}$ varies between $2.5$ and $15$. Fig. \ref{Fig4b} shows that $k_c\le 1$ when $X\_R_{system}\le 0.3$ and $k_c>1$ for  $X\_R_{system}>0.3$. Fig. \ref{Fig4b} also shows that $k_c$ is independent of $X\_R_{trx}$. Hence it can be expressed by a polynomial curve fit function of $X\_R_{system}$. A $4^{th}$ order polynomial is seen to be sufficient to approximate the correction factor $k_c$, with a goodness-of-fit, $R^2=0.995$, and is given by,
\begin{equation}\label{kc}
\begin{split}
k_c=&-0.011X\_R^{4}_{system}+0.112X\_R^{3}_{system}\\
&-0.348X\_R^{2}_{system}+0.564X\_R_{system}+0.884
\end{split}
\end{equation}
Fig. \ref{Fig4c} shows $\Delta J_{I\%,100ms}$ computed after including the $k_c$ factor for various $X\_R_{system}$ and $X\_R_{trx}$. The $\Delta J_{I\%,100ms}$ is found to be within the acceptable limit of $\pm 5\%$ which can be used to ensure that the prediction from the model will not result in erroneous MWT damage due to large error in $J_{I,t}$ calculations.
\begin{figure}[!t]
\centering
\includegraphics[keepaspectratio,scale=.43]{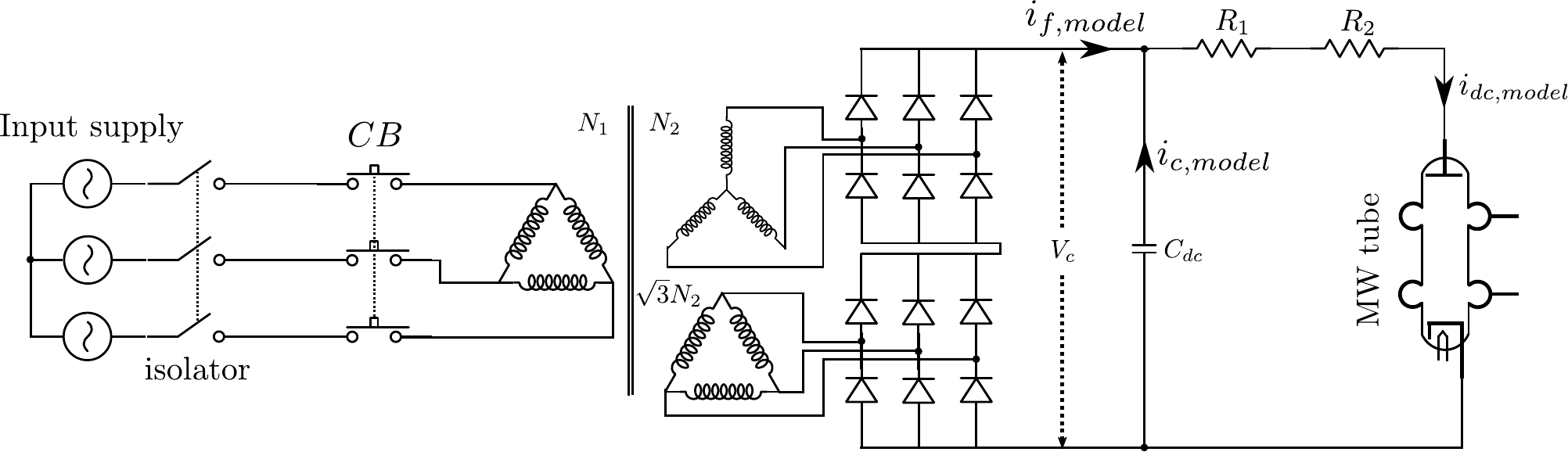}
\caption{Conventional 12-pulse HVPS feeding power to MWT and output of rectifiers are series connected.}
\label{HVPS12P_Sens_Series}
\end{figure}
\begin{figure}[!t]
\centering
\includegraphics[keepaspectratio,scale=0.34]{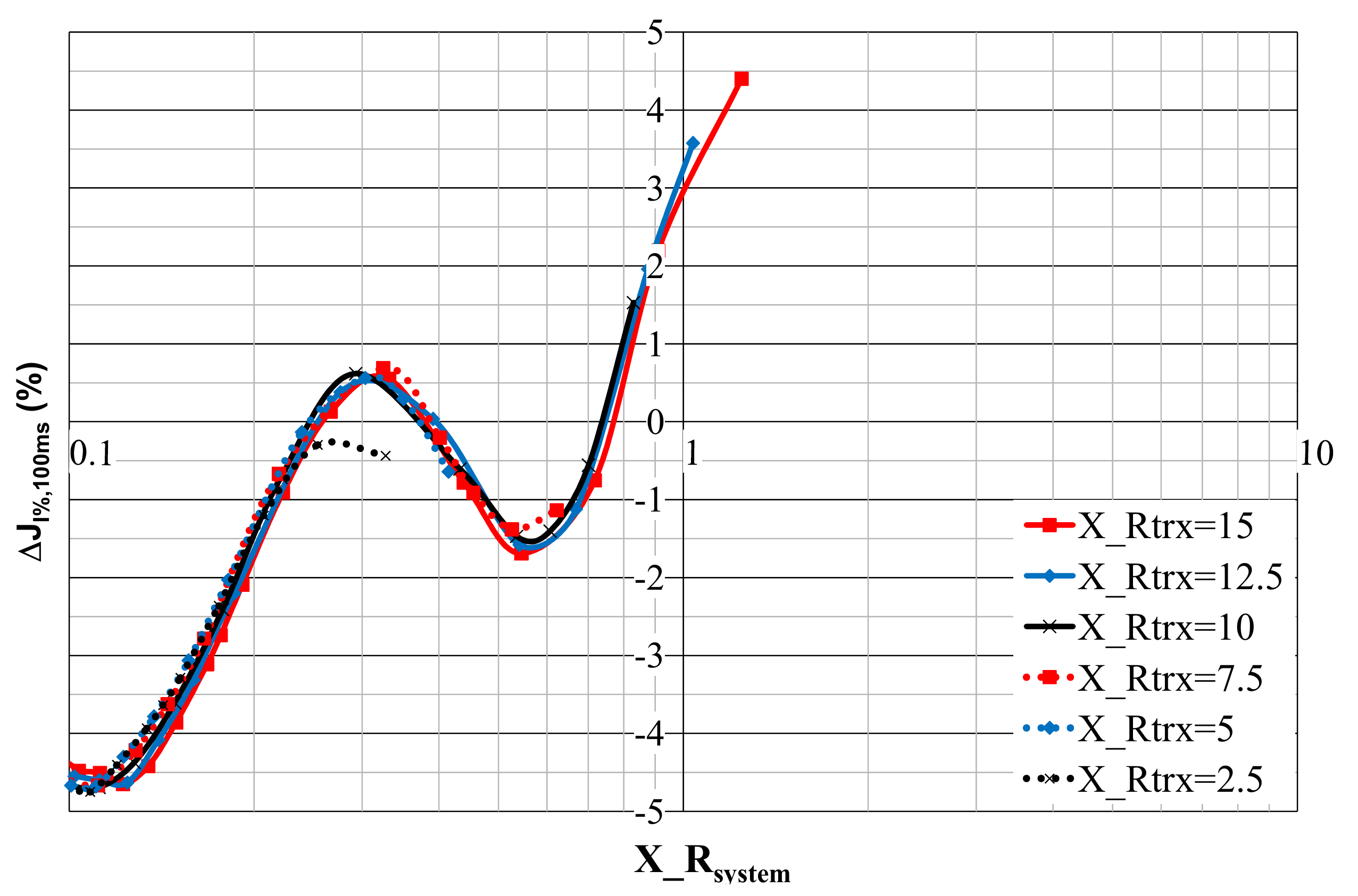}
\caption{$\Delta J_{I\%,100ms}$ for different $X\_R_{system}$ plotted for various $X\_R_{trx}$ after including $k_c$ factor for series connected dc output rectifier.}
\label{Fig4d}
\end{figure}
\subsection{Rectifier dc side is series connected}\label{3.4.4}
In some cases to achieve the required voltage the rectifier dc side is series connected as shown in Fig.~\ref{HVPS12P_Sens_Series}. For the parallel connected rectifier circuit, the impact of $R_L$ in the fault current is considered by transferring $R_L$ to the ac side using active power equality. Since the active power equality is same irrespective of the rectifier connection, the correction factor $k_c$ required for series or parallel connected connections are found to be closely matching. However due to the difference in the rectifier output connection the expression for base value of current $i_{f,base}$ in (\ref{ifaultdcMM}) and $R_{Lp}$ in (\ref{rlp}) should be modified. For both series and parallel connected rectifiers, keeping the same output power, the series connected rectifier input current is one half of that when compared to a parallel connected rectifier. Hence, the base value of current $i_{f,base}$ is given by,
\begin{equation}\label{ifaultdcMMS}
I_{f,base}=\dfrac{1}{2}\dfrac{\sqrt{2}E}{\sqrt{\left(R^{'}_{p}+k_cR_{Lp}\right)^{2}+X_{lp}^{'2}}}\dfrac{\sqrt{3}N_1}{N_2}k_{12}
\end{equation}
Since, for parallel connected rectifier circuit the expression for equivalent resistance and reactance referred to primary are derived by considering $R_L=0$, these expressions are also valid for series connected rectifier circuit. Hence, by applying (\ref{ifaultdcMMS}) in (\ref{AP}), the $R_{Lp}$ is given by,
\begin{equation}\label{rls}
R_{Lp}=\dfrac{1}{6}{\left(\dfrac{\sqrt{3}N_1}{N_2}k_{12}\right)}^2R_L
\end{equation}

The $\delta$ for series connected rectifier circuit can be obtained by substituting (\ref{rls}) and (\ref{Rtr}) in (\ref{deltaMM}), keeping the correction factor $k_c$ same as that derived in the previous section. The evaluated $\Delta J_{I\%,100ms}$ for the system with series connected rectifier is also in the range of $\pm 5\%$ using the correction factor $k_c$, shown in Fig.~\ref{Fig4d}.
\subsection{Complete model of dc fault current $i_{dc,model}(t)$}
For a given system $k_c$ is computed by applying (\ref{XRM}) into (\ref{kc}). The value of $k_c$ is used in (\ref{ifaultdcMM}) or (\ref{ifaultdcMMS}) to find $I_{f,base}$ based on whether the circuit is parallel connected or series connected at the output respectively. The evaluated $k_c$ is also used in (\ref{deltaMM}) to find $\delta$. The computed $I_{f,base}$ and  $\delta$ are used in (\ref{2ndOrdereqn}) and (\ref{if}) to obtain follow-on dc fault current model $i_{f,model}(t)$. The overall dc fault current can be found by superimposing $i_{f,model}(t)$ and $i_{c,model}(t)$ from (\ref{ic}) as,
\begin{equation}\label{totalf}
i_{dc,model}(t)=i_{f,model}(t)+i_{c,model}(t)
\end{equation}
This overall fault current model is used for design and evaluation of Joules Integral in the crowbar test circuit. The resulting error in the calculated $J_{I,t}$ from the model is expected to be less than $5\%$ as indicated by the analysis. This is validated using the experimental studies in section~\ref{exp}.
\begin{figure}[!t]
\centering
\includegraphics[keepaspectratio,scale=0.50]{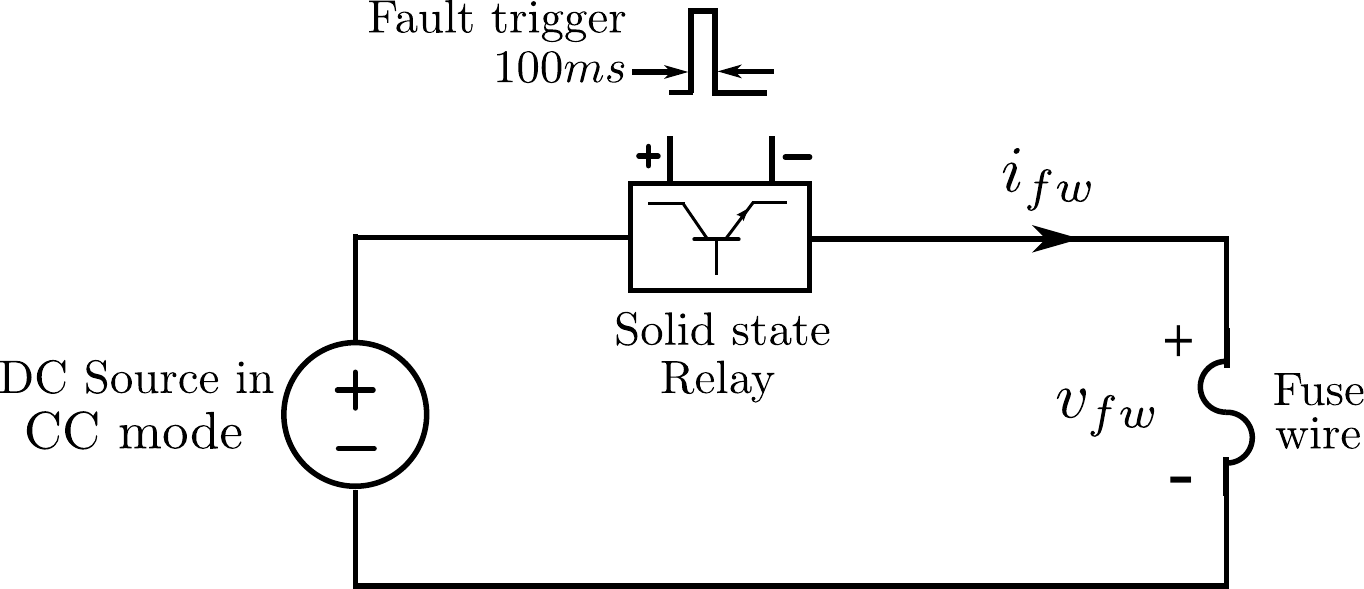}
\caption{Test circuit for the validation of fuse wire model.}
\label{FuseModel}
\end{figure}
\begin{figure}[!t]
\centering
\includegraphics[keepaspectratio,scale=0.38]{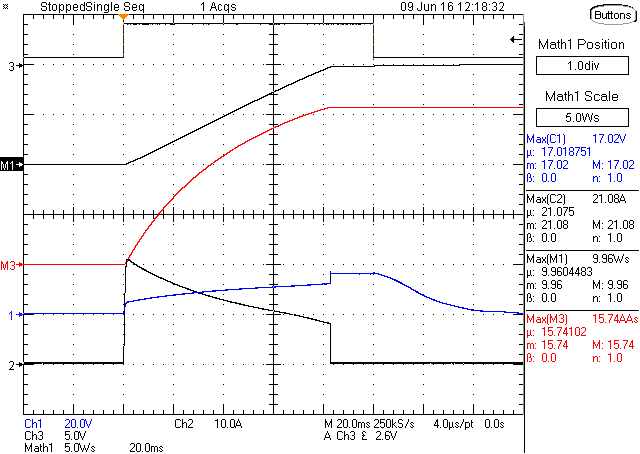}
\put(-190,155){\textsl{\small Fault trigger ($5V$/div)}}
\put(-233,134){\textsl{\normalsize Fuse energy ($5J$/div)}}
\put(-122,134){\textsl{\small Fuse $J_{I,t}$($5A^2s$/div)}}
\put(-120,73){\textsl{\normalsize $v_{fw}$ ($20V$/div)}}
\put(-110,38){\textsl{\normalsize $i_{fw}$ ($10A$/div)}}
\caption{Experimental waveforms of voltage, current, energy and $J_{I,t}$ of fuse wire from fuse wire test circuit. (Time scale: $20ms$/div).}
\label{FuseModelR}
\end{figure}
\section{Experimental results}\label{exp}
\subsection{Validation of MWT equivalent model}
A dc source in current limit mode is applied across a fuse wire for $100ms$ using a solid state relay (SSR) as shown in Fig. \ref{FuseModel}. The current limit is set for sufficient magnitude to melt the fuse wire close to $100ms$. This is to validate the assumption of neglecting the heat transfer due to conduction, convection and radiation in a small operational duration. From \cite{ref12} and \cite{Phil}, typical energy and Joules Integral requirement of a fuse wire for MWT application is $10J$ and $40A^2s$ respectively. Electrical conductivity of chosen fuse wire at room temperature is estimated experimentally and found to be $\sigma_o=5.13{\times}10^7S/m$. Other physical constants of copper is chosen as $\rho{=}8950kg/m^3$, $\mathcal{C}_p{=}395J/kg^0C$ and $\alpha_o{=}3.8{\times}10^{-3}/^0C$. The room and melting temperature of fuse wire is chosen as $T_o{=}30^0C$ and $1083^0C$ respectively.
\begin{figure}[!t]
\centering
\subfloat[]{\includegraphics[keepaspectratio,scale=0.60]{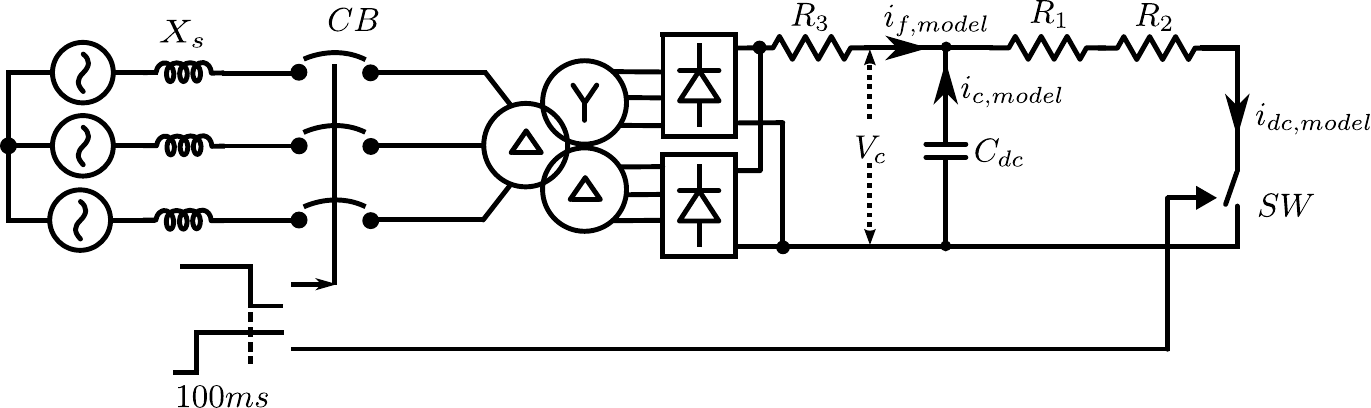}}\\
\subfloat[]{\includegraphics[keepaspectratio,scale=0.60]{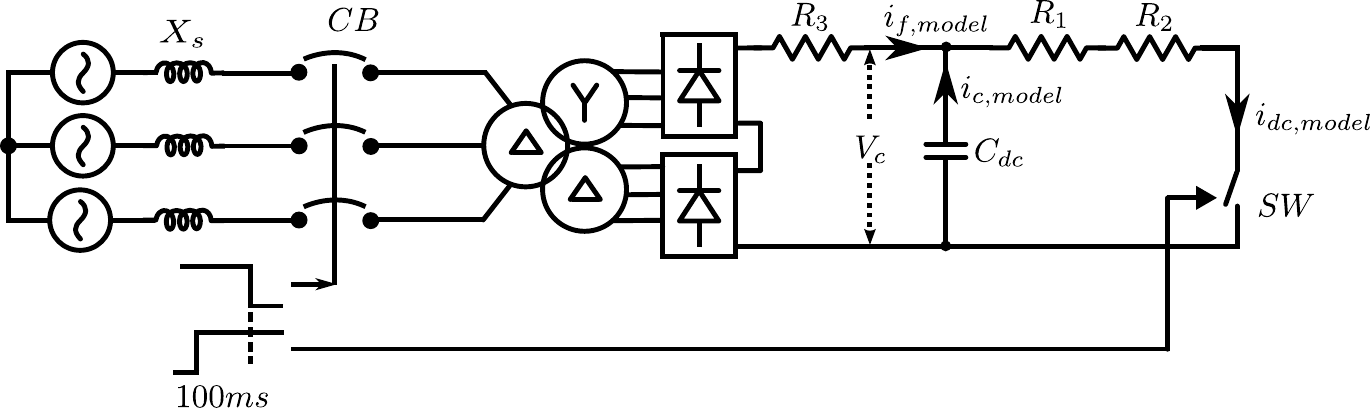}}
\caption{Test setup for validation of dc fault current (a) rectifiers are parallel connected (b) rectifiers are series connected.}
\label{Fig5add}
\end{figure}
\begin{table}[!t]
\renewcommand{\arraystretch}{1.3}
\caption{Parameters related to the test setup}
\label{parameter_testsetup}
\centering
\begin{tabular}{|l|l|r|}
\hline
\multicolumn{2}{|c|}{\bfseries Parameters} & \bfseries Values\\
\hline
&Primary&$\Delta$, $415V{+}15\%~l$-$l$ r.m.s., $50kVA$\\
&&$R_{p\Delta}{=}0.059\Omega$, $X_{l,p\Delta}{=}0.121\Omega$\\
\cline{2-3}
Transformer&Secondary&$\Delta$, $1100~l$-$l$ r.m.s., $25kVA$\\
\cline{3-3}
&&$Y$, $1100~l$-$l$ r.m.s., $25kVA$\\
\cline{3-3}
&&$R_{sp}{=}0.134\Omega$, $X_{l,sp}{=}0.209\Omega$\\
\hline
Resistances&$R_1$,$R_2$,$R_3$& $3\Omega$, $8\Omega$, $38\Omega$\\
\hline
Capacitance&$C_{dc}$&$92\mu F$\\
\hline
Input voltage&$E$&$465V$\\
\hline
Input frequency&$\omega$&$2\pi 50$~rad/s\\
\hline
Source impedance&$X_s$&$0.166\Omega$\\
\hline
\end{tabular}
\end{table}

Substituting the physical constants of copper in (\ref{JI_area}), to keep the $J_{Im}$ less than $40A^2s$, the $SWG$ of fuse wire should be less than $37$ and corresponds to diameter of $0.17mm$. Also from (\ref{energy4}), the product $Al$ required for a fuse wire of $10J$ is $2.6{\times}10^{-9}m^3$. Hence the length of fuse wire required for an energy of $10J$ is $110mm$. To avoid an arc through the air across the fuse wire, the fuse wire length is selected so that the length by voltage ratio is $10mm/kV$. Hence for a crowbar which operates at $12kV$, it is preferable to choose a fuse wire of higher length. The chosen length of fuse wire is equal to $165mm$. For constant $Al$ in (\ref{energy4}), the diameter required is $0.136mm$, equal to $39~SWG$ wire. Substituting the chosen length and diameter of $0.136mm$, in (\ref{JI_area}) and (\ref{energy4}) gives the $J_{Im}$ and $E_{fm}$ as $16.19A^2s$ and $9.51J$ respectively.

Fuse current and fuse voltage waveforms obtained from experiment given in Fig. \ref{FuseModelR} shows that fuse melting took place at $84ms$. The $J_{Im}$ in the fuse wire at the time of melting is recorded from the experiment as $15.7A^2s$ and is shown in Fig. \ref{FuseModelR}. Also from Fig. \ref{FuseModelR} the energy in the fuse wire at the time of melting is $9.96J$. The $J_{Im}$ and $E_{fm}$ of the fuse wire from the model is obtained as $16.19A^2s$ and $9.51J$ respectively. The closely matching experimental results with the computed value establish the assumption of insignificant contribution of heat transfer due to conduction, convection and radiation for crowbar applications.
\begin{figure}[!t]
\centering
\includegraphics[keepaspectratio,scale=0.38]{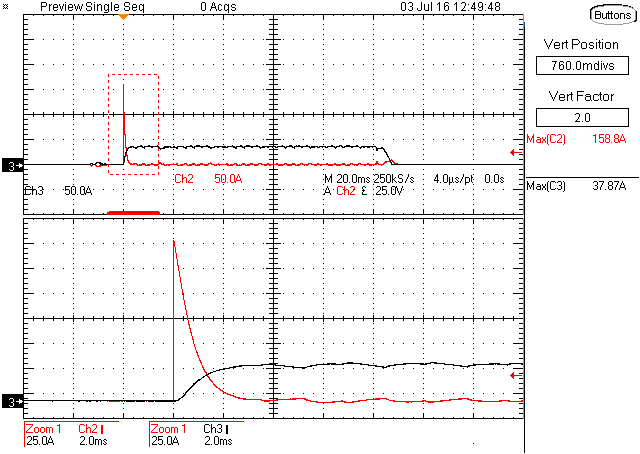}
\put(-169,76){\textsl{\normalsize Capacitor discharge current}}
\put(-169,63){\textsl{\normalsize ($25A$/div)}}
\put(-153,38){\textsl{\small Follow-on current ($25A$/div)}}\hspace{0.025in}
\caption{DC capacitor discharge and follow-on current (Time scale: $20ms$/div, Zoom: $2ms$/div).}
\label{Fig6b}
\end{figure}
\begin{figure}[!t]
\centering
\includegraphics[keepaspectratio,scale=0.38]{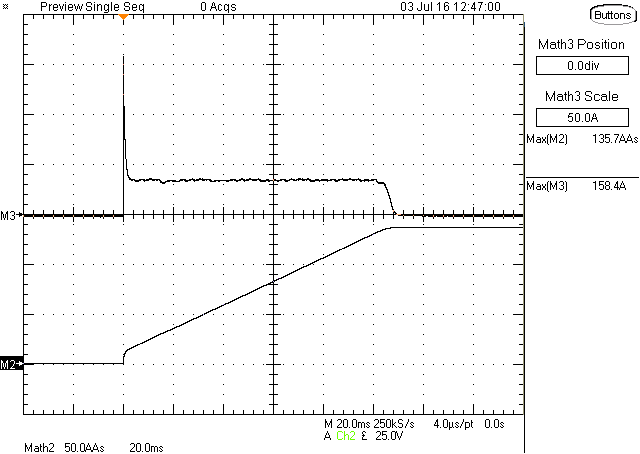}
\put(-194,119){\textsl{\normalsize Fuse current ($50A$/div)}}
\put(-230,65){\textsl{\normalsize Fuse $J_{I,t}$($50A^2s$/div)}}
\caption{Fuse current and its $J_{I,t}$ for the parallel connected rectifier test circuit (Time scale: $20ms$/div).}
   \label{Fig6c}
\end{figure}
\subsection{Validation of dc fault current model $i_{dc,model}(t)$}
The test circuit used for the validation of dc fault current is shown in Fig. \ref{Fig5add}(a) and (b) for parallel connected rectifier circuit and series connected rectifier circuit respectively. The approach of adding the fault contributions of the two equivalent circuit models for dc fault current shown in Figs. \ref{Fig2}(a) and (b) can be established if the summation in (\ref{totalf}) can be shown to be accurate. This should happen even when the resistance involved in the capacitor discharge ($i_{c,model}(t)$) is different from the resistance involved in calculating $i_{f,model}(t)$ as can be observed in Fig.~\ref{Fig5add}(a) and (b). The resistance $R_3$ added between the diode bridge and $C_{dc}$ in Fig. \ref{Fig5add}(a) and (b), which limits the follow-on current, giving $R_L=R_1+R_2+R_3$ for (\ref{rlp}) where as $R_L=R_1+R_2$ is for the dc capacitor discharge model in (\ref{ic}). Other parameters related to the test circuit in Fig. \ref{Fig5add}(a) and (b) and required to find the dc fault current model are given in Table~\ref{parameter_testsetup}. Fig. \ref{Fig6b} waveforms show the follow-on current and dc capacitor discharge current from experiments. It can be observed that in one time constant of the dc capacitor discharge current there is no significant contribution of the follow-on current in terms of $J_{I,t}$. This confirms the assumption of the ability to sum the fault contributions of the two equivalent circuit models.
\subsubsection{Parallel connected rectifier output}
Using Table~\ref{parameter_testsetup} in (\ref{XRM}) and in (\ref{kc}) the $X\_R_{system}$ and $k_c$ are computed as $0.052$ and $0.912$ respectively. The dc fault current model for the circuit in Fig. \ref{Fig5add}(a) precharged to $V_c=1700V$ and having parameters in Table~\ref{parameter_testsetup} is given by,
\begin{equation}\label{modelp}
\begin{split}
i_{f,model}(t)=&33.77\big(1-e^{-5454.6 t}\big(\cos 334.2t\\
&+16.32\sin 334.2t\big)\big)+154.54e^{-988.14t}
\end{split}
\end{equation}
The Joules Integral $J_{I,t}$ of (\ref{modelp}) computed at time $t=104ms$, denoted as $J_{I,104ms}$, is $137.7A^2s$. Fig. \ref{Fig6c} shows the experimental results of dc fault current and $J_{I,t}$ for the case of parallel connected rectifier. From experiment $J_{I,104ms}$ is found to be $135.7A^2s$. The $\Delta J_{I\%,104ms}$ and the peak of fault current are found to be within the acceptable limit of $\pm 5\%$.
\begin{figure}[!t]
\centering
\includegraphics[keepaspectratio,scale=0.38]{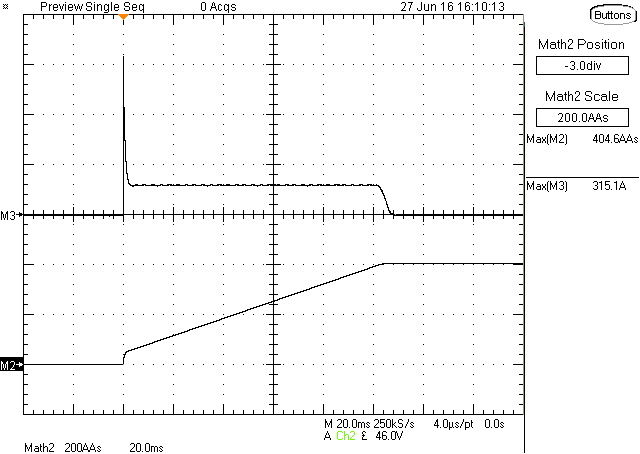}
\put(-195,119){\textsl{\normalsize Fuse current ($100A$/div)}}
\put(-225,65){\textsl{\normalsize Fuse $J_{I,t}$($200A^2s$/div)}}
\caption{Fuse current and its $J_{I,t}$ for the series connected rectifier test circuit (Time scale: $20ms$/div).}
   \label{Fig7}
\end{figure}
\begin{table}[!t]
\renewcommand{\arraystretch}{1.3}
\caption{Summary of results from the MWT fault model and the experimental test setup}
\label{testresults}
\centering
\begin{tabular}{|l|l|r|r|r|}
\hline
\bfseries Test circuit & \bfseries Parameters&\multicolumn{2}{c|}{\bfseries From}&\bfseries Error\\
\cline{3-4}
&&\bfseries Model& \bfseries Experiment &(\%)\\
\hline
Rectifier in&$J_{I,104ms}~(A^2s)$&$137.70$&$135.70$&$-1.47$\\
\cline{2-5}
Fig. \ref{Fig5add}(a)&Peak current~(A)&$154.54$&$158.40$&$2.44$\\
\hline
Rectifier in&$J_{I,102ms}~(A^2s)$&$413.10$&$404.60$&$-2.10$\\
\cline{2-5}
Fig. \ref{Fig5add}(b)&Peak current~(A)&$309.09$&$315.10$&$1.91$\\
\hline
Fuse wire&$J_I~(A^2s)$&$16.19$&$15.70$&$3.00$\\
\cline{2-5}
&Energy~(J)&$9.51$&$9.96$&$-4.70$\\
\hline
\end{tabular}
\end{table}
\subsubsection{Series connected rectifier output}
An experiment is also carried out with series connection at the rectifier output of Fig. \ref{Fig5add}(b) keeping all other parameters same as in Table~\ref{parameter_testsetup}. The $X\_R_{system}$ and $k_c$ computed for series connection are $0.2048$ and $0.9858$ respectively. For series connection and $V_c=3400V$, the dc fault current model is given by,
\begin{equation}\label{models}
\begin{split}
i_{f,model}(t)=&59.67\big(1-e^{-1513.6 t}\big(\cos 334.2t\\
&+4.53\sin 334.2t\big)\big)+309.09e^{-988.14t}
\end{split}
\end{equation}
The Joules Integral $J_{I,t}$ of (\ref{models}) computed at time $t=102ms$, denoted as $J_{I,102ms}$, is $413.1A^2s$. Fig. \ref{Fig7} shows the dc fault current and $J_{I,102ms}$ obtained experimentally for series connection, where $J_{I,102ms}$ is obtained to be $404.6A^2s$. For parallel and series connection the error in $J_{I,t}$ and the peak value of fault current are found to be well within an acceptable limit of $\pm 5\%$. Various results obtained from MWT model and dc fault current model are summarized in Table \ref{testresults}.
Test results of $10kV$, $1kA$ crowbar built based on the model to limit the energy in MWT below $10J$ are given in~\cite{ref15n}.
\section{Conclusion}
The MWT used in many applications demands protection against the excess energy released into the tube during the internal arc fault. For the proper design of a protective crowbar device, computation of energy into the MWT is essential. With the knowledge of the model of dc fault current and MWT model the energy into MWT can be computed. The paper discusses a simplified model for the MWT as an equivalent fuse wire for crowbar applications. The paper presents a dc fault-current model based on Joules Integral energy concept. The model utilizes the summation of fault contribution of capacitor discharge and follow-on current that are treated as independent circuits. The power balance approach is proposed  to transform resistance from dc side to ac side and a correction factor is applied on the transformed resistance to bring the error in $J_{I,t}$ less than $\pm 5\%$. This model also provides flexibility to use different values of $X/R$ ratio of transformer and also allows the use of a range of dc current limiting resistances used in the test circuit. The model is validated for both series and parallel connection of ac-dc $12$ pulse rectifier bridges. The paper also provides a simplified model for the MWT. All the analytical results are found to be closely matching with the experimental results, with errors less than $5\%$. Using the model a crowbar of $10kV$ and $1kA$ is also built to limit the energy in MWT below $10J$.


\section*{Acknowledgment}
The work is supported by Ministry of Electronics and Information Technology, Govt. of India, through NaMPET programme and Department of Atomic Energy (DAE), Government of India through Institute for Plasma Research, Gandhinagar, India.



%
%
%

\bibliographystyle{IEEEtran}
\bibliography{Reference}

\begin{thebibliography}{10}
\providecommand{\url}[1]{#1}
\csname url@samestyle\endcsname
\providecommand{\newblock}{\relax}
\providecommand{\bibinfo}[2]{#2}
\providecommand{\BIBentrySTDinterwordspacing}{\spaceskip=0pt\relax}
\providecommand{\BIBentryALTinterwordstretchfactor}{4}
\providecommand{\BIBentryALTinterwordspacing}{\spaceskip=\fontdimen2\font plus
\BIBentryALTinterwordstretchfactor\fontdimen3\font minus
  \fontdimen4\font\relax}
\providecommand{\BIBforeignlanguage}[2]{{%
\expandafter\ifx\csname l@#1\endcsname\relax
\typeout{** WARNING: IEEEtran.bst: No hyphenation pattern has been}%
\typeout{** loaded for the language `#1'. Using the pattern for}%
\typeout{** the default language instead.}%
\else
\language=\csname l@#1\endcsname
\fi
#2}}
\providecommand{\BIBdecl}{\relax}
\BIBdecl

\bibitem{ref1}
M.~Gilmore, ``Engineering applications of plasma science,'' \emph{IEEE
  Potentials}, vol.~17, no.~3, pp. 4--8, Aug. 1998.

\bibitem{ref2}
Y.~J. Kim, S.~Jin, G.~H. Han, G.~C. Kwon, J.~J. Choi, E.~H. Choi, H.~S. Uhm,
  and G.~Cho, ``Plasma apparatuses for biomedical applications,'' \emph{IEEE
  Trans. Plasma Science}, vol.~4, no.~43, pp. 944--950, Apr. 2015.

\bibitem{ref3}
D.~Leonhardt, C.~Muratore, and S.~G. Walton, ``Applications of electron-beam
  generated plasmas to materials processing,'' \emph{IEEE Trans. Plasma
  Science}, vol.~2, no.~33, pp. 783 -- 790, Apr. 2005.

\bibitem{ref4}
X.~Denga, A.~Nikiforova, C.~Leys, D.~Vujosevicb, and V.~Vuksanovic,
  ``Preparation of antibacterial non-woven fabric via atmospheric pressure
  plasma process,'' in \emph{Proc. IEEE Int. Conf. Plasma Sci.}, Antalya, May
  2015, p.~1.

\bibitem{ref5}
L.~Popelier, A.~Aanesland, S.~Mazouffre, and P.~Chabert, ``Extraction and
  acceleration of ions from an ion-ion plasma-application to space
  propulsion,'' in \emph{5th Int. Conf. Recent Advances in Space Technologies
  (RAST)}, Istanbul, June 2011, pp. 708--711.

\bibitem{ref6}
{S.D. Korovin, V.V. Rostov, S.D. Polevin, I.V. Pegel, E. Schamiloglu, M.I.
  Fuks, and R.J. Barker}, ``Pulsed power-driven high-power microwave sources,''
  \emph{Proc. of the IEEE}, vol.~92, no.~7, pp. 1082--1095, July. 2004.

\bibitem{ref7}
D.~Yaogen, P.~Jun, N.~Delu, Z.~Shichang, F.~Chunjiu, L.~Tieshan, Z.~Yunsu, and
  L.~Jiron, ``Research progress and advances on high power klystron,'' in
  \emph{Proc. Int. Conf. Electronics and Radiophysics of Ultra-High
  Frequencies}, St Petersburg, Aug. 1999, pp. 32 -- 36.

\bibitem{ref11}
S.~G.~E. Pronko and T.~E. Harris, ``A new crowbar system for the protection of
  high power gridded tubes and microwave devices,'' in \emph{Proc. IEEE Int.
  Vaccum Electronics Conf. (IVEC)}, Noordwijk, April 2001.

\bibitem{ref15}
T.~G. Subhash~Joshi and V.~John, ``Design and evaluation of mounting clamps for
  crowbar application,'' in \emph{Proc. National Power Electronics Conference},
  Bombay, Dec. 2015, pp. 1--6.

\bibitem{ref8}
P.~Pozzobon, ``Transient and steady-state short-circuit currents in rectifiers
  for dc traction supply,'' \emph{IEEE Trans. Vehicular Technology}, vol.~47,
  no.~4, pp. 1390--1404, Nov. 1998.

\bibitem{ref9_}
S.~Chiniforoosh, H.~Atighechi, A.~Davoudi, J.~Jatskevich, A.~Yazdani,
  S.~Filizadeh, M.~Saeedifard, J.~A. Martinez, V.~Sood, K.~Strunz,
  J.~Mahseredjian, and V.~Dinavahi, ``Dynamic average modeling of front-end
  diode rectifier loads considering discontinuous conduction mode and
  unbalanced operation,'' \emph{IEEE Trans. Power Delivery}, vol. 275, no.~1,
  pp. 421--429, Jan. 2012.

\bibitem{ref10}
K.~L. Lian, B.~K. Perkins, and P.~W. Lehn, ``Harmonic analysis of a three-phase
  diode bridge rectifier based on sampled-data model,'' \emph{IEEE Trans. Power
  Delivery}, vol.~23, no.~2, pp. 1088--1096, Apr. 2008.

\bibitem{ref15_}
T.~G. Subhash~Joshi and V.~John, ``{Microwave Tube Fault-Current Model for
  Design of Crowbar Protection},'' in \emph{IEEE Int. Conf. on Power
  Electronics, Drives and Energy Systems (PEDES)}, Trivandrum, Dec. 2016, pp.
  1--6.

\bibitem{ref12}
{Y.S.S. Srinivas, M. Kushwah, S.V. Kulkarni, K. Sathyanarayana, P.L. Khilar, B.
  Pal, P. Shah, A.R. Makwana, B.R. Kadia, K.M. Pannar, S. Dani, R. Singh, K.G.
  Pannar and D. Bora}, ``{Results of 10-Joule wire-burn test performed on 70kV
  rail-gap crowbar protection system for high power klystrons and gyrotron},''
  \emph{19th Symposium on Fusion Engineering}, pp. 91--94, 2002.

\bibitem{ref13}
M.~S. Agarwal, A.~D. Stokes, and P.~Kovitya, ``Pre-arcing behaviour of open
  fuse wire,'' \emph{Journal of Physics D: Applied Physics}, vol.~20, no.~10,
  pp. 1237--1242, Oct. 1987.

\bibitem{ref14}
K.~C. Chen, L.~K. Warne, Y.~T. Lin, R.~L. Kinzel, J.~D. Huﬀ, M.~B. McLean,
  M.~W. Jenkins, and B.~M. Rutherford, ``Conductor fusing and gapping for bond
  wires,'' \emph{Progress In Electromagnetics Research M}, vol.~31, pp.
  199--214, May 2013.

\bibitem{Phil}
{Philips GmbH}, ``{Installing and Operating Klystron YK1360},'' PHILIPS manual
  No.1, Oct. 1992.

\bibitem{ref15n}
T.~G. Subhash~Joshi and V.~John, ``{Performance Comparison of ETT- and
  LTT-Based Pulse Power Crowbar Switch},'' \emph{IEEE Trans. Plasma Science},
  pp. 1--7, {to be published [DOI:10.1109/TPS.2017.2759668]}.

\end{thebibliography}

\end{document}